\documentclass[conference]{IEEEtran}
\usepackage{cite}
\usepackage{amsmath,amssymb,amsfonts}
\usepackage{algorithmic}
\usepackage{graphicx}
\usepackage{textcomp}
\usepackage{xcolor}
\usepackage{cite}
\usepackage{amsmath,amssymb,amsfonts}
\usepackage{algorithmic}
\usepackage{graphicx}
\usepackage{textcomp}
\usepackage{xcolor}
\usepackage{subcaption}
\usepackage{multicol,multirow}
\usepackage{enumerate,textcomp,array}
\usepackage{placeins}
\usepackage{nicematrix}
\NiceMatrixOptions{cell-space-top-limit = 1pt,cell-space-bottom-limit = 1pt}
\usepackage{threeparttable}
\usepackage{dblfloatfix}
\usepackage{comment}
\usepackage{array}
\newcolumntype{?}{!{\vrule width 1pt}}
\usepackage{gensymb}

\usepackage{filecontents}
\usepackage{tikz, tikzscale, pgfplots}
\usetikzlibrary{calc}

\newcommand{\R}{\mathbb{R}}
\newcommand{\rmd}{\mathrm{d}}

\newcommand{\bA}{\mathbf{A}}
\newcommand{\bB}{\mathbf{B}}
\newcommand{\bD}{\mathbf{D}}
\newcommand{\bgam}{\mathbf{\gamma}}
\newcommand{\bK}{\mathbf{K}}

\newcommand{\bone}{\mathbf{1}}
\newcommand{\bM}{\mathbf{M}}
\newcommand{\bT}{\mathbf{T}}
\newcommand{\bzero}{\mathbf{0}}

\newcommand{\tbB}{\tilde{\mathbf{B}}}

\newcommand{\cG}{\mathcal{G}}
\newcommand{\cI}{\mathcal{I}}

\newcommand{\diag}{\mathrm{diag}}

\usepackage{url}

\usepackage{breakurl}
\usepackage{booktabs}

\def\BibTeX{{\rm B\kern-.05em{\sc i\kern-.025em b}\kern-.08em
    T\kern-.1667em\lower.7ex\hbox{E}\kern-.125emX}}
\begin{document}

\title{Computationally Efficient Analytical Models of Frequency and Voltage in 
Low-Inertia Systems\\}
\author{\IEEEauthorblockN{Marena Trujillo\IEEEauthorrefmark{1}\IEEEauthorrefmark{2}\IEEEauthorrefmark{3}, Amir Sajadi \IEEEauthorrefmark{2}\IEEEauthorrefmark{3}, Jonathan Shaw\IEEEauthorrefmark{4}, and Bri-Mathias Hodge\IEEEauthorrefmark{1}\IEEEauthorrefmark{2}\IEEEauthorrefmark{3}\IEEEauthorrefmark{4}}
	\IEEEauthorblockA{\IEEEauthorrefmark{1}Department of Electrical, Computer, and Energy Engineering, University of Colorado of Boulder,
		Boulder, Colorado 80309}
	\IEEEauthorblockA{\IEEEauthorrefmark{2}Renewable and Sustainable Energy Institute, University of Colorado of Boulder,
		Boulder, Colorado 80309}
	\IEEEauthorblockA{\IEEEauthorrefmark{3}National Renewable Energy Laboratory, 
		Golden, Colorado 80401}
	\IEEEauthorblockA{\IEEEauthorrefmark{4}Department of Applied Mathematics, University of Colorado of Boulder,
		Boulder, Colorado 80309\\
		\{marena.trujillo, amir.sajadi, jonathan.shaw, brimathias.hodge\}@colorado.edu}
}

\maketitle
\thispagestyle{empty}
\pagestyle{plain}
\begin{abstract}
In this paper, low-order models of the frequency and voltage response of mixed-generation, low-inertia systems are presented. These models are unique in their ability to efficiently and accurately model frequency and voltage dynamics without increasing computational burden as the share of inverters increases in a system. The models are validated against industry-grade electromagnetic transient simulation, compared to which the proposed models are several orders of magnitude faster. The accuracy and efficiency of the low-inertia frequency and voltage models makes them well suited for a variety of planning and operational studies, especially for multi-scenario and probabilistic studies, as well as for screening studies to establish impact zones based on the dynamic interactions between inverters and synchronous generators.
\end{abstract}

\begin{IEEEkeywords}
Frequency response, Low-inertia systems, Grid-forming inverters, Voltage dynamics
\end{IEEEkeywords}
%
%
\section{Introduction}\label{intro}
The decarbonization of power sectors worldwide is a crucial aspect of efforts to limit carbon emissions and mitigate the worst effects of a changing climate \cite{ting2022mitigating}. Among the carbon-free sources of electricity generation, solar photovoltaics and wind power are cost-competitive in most locations compared to legacy fossil fuel generation \cite{2024LCOEReport}, which has spurred their widespread deployment. These clean energy technologies are inverter-based resources (IBRs), which means they interact with the larger power system through a power electronics interface. Conventional generation units, such as those powered by nuclear, natural gas, and coal, are equipped with synchronous generators (SGs), which were invented almost a century ago and are electromechanically coupled to the rest of the power system \cite{sauerPowerSystemDynamics}. Notably, these two distinct classes of electricity generation, IBRs and SGs, operate on substantially different principles and, as a result, the mathematical models representing them are also vastly different.

Given the long history of SGs, there are several effective ways to simulate large SG-dominated systems. Electromagnetic transient (EMT) simulations can provide highly accurate results by using high-order device models and dynamic line models, but simulations of large-scale systems are infeasible due to the degree of computational complexity \cite{kenyonComparisonofElectromagnetic}. Combining IBRs and SGs in large EMT models can introduce even more computational difficulties due to the different timescales of their dynamics. Solving such a numerically stiff system with an explicit fixed time step method requires very small time steps to maintain numerical stability, which increases the computational burden significantly \cite{brenan1995numerical}. Phasor domain models are much faster than EMT simulation, but they are intended for studying low-frequency phenomena and assume near
constant system frequency. Consequently, they are unsuitable for modeling the fast dynamics introduced by IBRs \cite{kenyonComparisonofElectromagnetic,laraRevisiting}. System frequency models (SFRs) take order-reduction to the extreme by neglecting all network effects and representing the frequency response of the entire system as that of a single device \cite{AndersonALowOrderSystemFrequency,chanDynamicEquivalentsAverage1972}. While SFRs can be useful for modeling small, coherent clusters, these models are ill-suited for representing systems with both IBRs and SGs, and they provide no information on voltage dynamics. Thus, there is a pressing need for models that can efficiently and accurately simulate both the frequency and voltage dynamics of large-scale modern power systems with high shares of IBRs.

The contribution of this paper is a novel reduced-order mathematical formulation of frequency and voltage dynamics in a mixed-generation system that allows for the exceedingly fast simulation of system dynamics with a timestep in the $\mu s$ range.
The proposed models were extensively validated against high-order EMT simulations, compared to which they demonstrated promising accuracy and solve times orders of magnitude faster. Unlike in EMT simulation, where the degree of computational difficulty increases with the share of IBRs in a system, the solve times of proposed models \textit{accelerate} as the proportion of IBRs to SGs increases. The interpretability, speed, and linearity of the models makes them attractive for numerous applications, including as an EMT-screening tool and for integration into an optimization framework for the formulation of stability constraints.

\section{Comparison of SG and IBR Dynamic Behavior}\label{motivation}
The frequency and voltage dynamics of IBRs differ substantially from those of SGs \cite{sajadiSynchronizationElectricPower2022}. As a result, their interactions in a system with a significant shares of both technologies could give rise to emerging dynamics, which may manifest themselves in the form of new transient phenomena not conventionally observed in SG-dominated systems. Examples of some of these interactive dynamics have been studied in the recent literature \cite{hatziargyriouDefinitionClassificationPower2021,markovic2021understanding} and have been observed in industrial practice \cite{EssentialActionsIndustry2025,InverterBasedResourceStrategy}. Here, we discuss the specific ways IBR control schemes could introduce new dynamics. In this section, we make the distinction between grid-forming inverters (GFMs) and grid-following inverters (GFLs). Within the GFM category, we further differentiate between GFMs with a multi-loop droop control scheme \cite{LasseterGrid-FormingInverters} and those with a virtual synchronous generator (VSG) control scheme \cite{DriesenVSG}. When not specified, ``GFM" refers to a multi-loop droop GFM, while ``VSG" is used for virtual synchronous generators. Finally, distinctions are made between GFLs with and without grid-support functionality. 

\subsection{SG vs. IBR design basics}

In the North American grid AC power is generated at 60 Hz and voltage is maintained within certain limits, depending on the specific location in the system. Generation devices that are capable of independently forming voltage are \textit{grid-forming resources}. This is a more broad term than GFM; in fact, SGs are grid-forming resources, because they form their own frequency and voltage. In a SG, the rapid process of regulating the voltage magnitude at the point of interconnection is only loosely coupled with the frequency regulation process, which occurs on a slower timescale \cite{sauerPowerSystemDynamics}. GFMs are also grid-forming resources, meaning they too act as voltage sources \cite{linResearchRoadmapGridForming}. However, GFLs act as current, not voltage, sources and are therefore not grid-forming resources\cite{LiRevisiting}. Instead of actively managing their voltage and frequency, they regulate their active and reactive power injections and rely on measurements from an internal phase-locked loop (PLL) to synchronize with the rest of the grid.

\subsection{Frequency dynamics}
When a disturbance occurs in a power system that results in a power imbalance, the generators connected to the system must correct the resultant imbalance to prevent frequency deviations. Consider a power generation contingency event, when there is a sudden drop in generated active power and reactive power, e.g., a generator unit trip. In a system of SGs, this power imbalance is initially addressed by the kinetic energy of the rotating mass of each SG, which slows down that frequency deviation just long enough for the controller to activate \cite{machowskiPowerSystemDynamics2020}. In a sense, in AC power networks, the synchronized frequency acts as a communication signal whose deviations can be detected by all equipment and trigger corrective or preventative actions. The first group of controllers that take action within the first few seconds after a frequency deviation is propagated through the network are the turbine governors to arrest further decline in frequency by adjusting the active power output of SGs. Seconds after their action, most commonly, the system frequency settles at a new lower level. Collectively, these actions make up the frequency response of SGs \cite{machowskiPowerSystemDynamics2020}. Importantly, this process is electro-mechanical in nature since the frequency of the voltage phasor is constructed by the mechanical rotation of the machine's rotor and is proportional to the speed at which the machine rotates. When there is a mismatch between the mechanical input power and electrical output power, the machine either speeds up or down and, subsequently, frequency either increases or decreases. As a result of this coupling, the inertia of the generator rotor dictates the timescale of frequency dynamics \cite{machowskiPowerSystemDynamics2020}. 

This process of frequency regulation in GFM is quite different from that in SG, mainly due to the absence of electromechanical elements, and instead the central role of control algorithms that govern the operation of the inverter. There are several control paradigms that dictate frequency response in GFMs. A multi-loop droop GFM rapidly adjusts the frequency setpoint of the inverter to account for a change in active power output \cite{LasseterGrid-FormingInverters}. This change in frequency happens at a much faster timescale (typically $< 1$s) than in a SG (typically $>10$s). VSGs, on the other hand, are designed to mimic the frequency dynamics of synchronous generators \cite{beckVSMS}. In other words, the speed of their frequency response is not physically limited by the inertia of a rotating mass as it is in a SG, rather it is artificially slowed down as to not introduce dynamics that are atypical of SG-dominated systems. On the seconds timescale, the frequency response of SGs and VSGs is second-order in nature, while multi-loop GFMs is first-order \cite{sajadiSynchronizationElectricPower2022}. GFLs with grid-support functionality may contribute to frequency response when weak grid conditions are not present. GFLs without grid-support functionality do not contribute towards correcting frequency deviations.

\subsection{Voltage dynamics}
In a SG, voltage control is achieved through the excitation system, which consists of an exciter and an automatic voltage regulator (AVR). An exciter provides DC current to the rotor windings of an SG and the resultant magnetic field within the generator induces an electromotive force (emf) in each of the stator windings. If the terminal voltage of a generator drops due to a disturbance, the AVR uses measurements of the generator terminal voltage to produce a voltage error signal which is used to alter the output of the exciter and correct the voltage error \cite{machowskiPowerSystemDynamics2020}. While the AVR does not explicitly use reactive power to regulate voltage, the terminal voltage of a generator is directly related to its reactive power output. Consider a one machine, infinite bus system. Assuming the system is sufficiently stiff and the power angle is small, 
the sensitivity of reactive power injection by the machine could be approximated by \eqref{Q_Ef} \cite{machowskiPowerSystemDynamics2020}. 
\begin{equation}\label{Q_Ef}
	\Delta Q \approx x_d^{-1}\cdot \Delta E_f 
\end{equation}  
where $E_f$ is the internal emf and $x_d$ is the synchronous reactance. In a SG, reactive power injection and absorption are capped by the field heating and stator end heating limits, respectively. In practice, however, the power rating of an exciter tends to be quite large, and enough reactive power can be supplied so that voltage recovers quickly, before frequency dynamics occur \cite{machowskiPowerSystemDynamics2020}. Active power is constrained by the turbine power limits. In this way, the active and reactive power limitations of an SG are only loosely related \cite{machowskiPowerSystemDynamics2020}. If the generator is unable to meet these reactive power demands due to limits, then the AVR will be unable to restore the generator's terminal voltage to its pre-disturbance level. This voltage regulation is an electro-magnetic process, and it occurs on a faster timescale than the frequency response of a SG.

In a multi-loop GFM, the dc-link and the Q-V droop are analogous to the excitation system of an SG \cite{NasrPowerSwing,xuStabilityDC}. However, unlike in SGs, where post-disturbance voltage dynamics are driven by a voltage error signal, a reactive power error signal serves as the input to Q-V droop control. The reactive power output of the inverter is used to control the terminal voltage \cite{rocabert2012control}. Because a GFM is not calculating a voltage error based on its terminal voltage and a pre-disturbance setpoint, the terminal voltage is not restored to its original value, as it is in a SG. Rather, an new voltage setpoint is created, leading to a markedly different voltage response shape as compared to that seen in a SG. It is also common for VSGs to employ Q-V droop control to regulate terminal voltage \cite{cheema2020comprehensive}.
\section{Frequency Model}\label{fmodel}
The unique computational efficiency of the proposed frequency model is achieved through the use of a Kron-reduced network model that is predicated on DC power flow assumptions and low-order GFM and SG frequency response models that are decoupled from reactive power/voltage device dynamics. Here, we first introduce the linearized network model before presenting the frequency models for individual generating devices. The network model and device models are then combined to formulate the low-inertia frequency response model.

\subsection{Network model}\label{network}
In \cite{trujillo2025analyticalmodelsfrequencyvoltage}, the authors demonstrate that reactive power-voltage dynamics can be reasonably neglected when modeling frequency response. Here we utilize the same network model, given in (\ref{delta_p_gen}), which is based on DC power flow approximations:
\begin{equation}\label{networkeqn}
	\begin{bmatrix} \Delta P_{G} \\ \Delta P_{L}\end{bmatrix} =  \begin{bmatrix} \bB_{GG} & \bB_{GL} \\ \bB_{LG} & \bB_{LL} \end{bmatrix} \begin{bmatrix} \Delta \delta_{G} \\ \Delta \delta_{L}\end{bmatrix}
\end{equation}
where a $G$ subscript indicates a generator bus and an $L$ subscript indicates a non-generator bus. The network is topologically reduced through the elimination of non-generator nodes \cite{machowskiPowerSystemDynamics2020}, as follows:
\begin{equation}\label{delta_p_gen}
	\Delta P_G = \bB_r \Delta \delta_G + \bB_L \Delta P_{L}, 
\end{equation}
where $\bB_{r}= \bB_{GG}-\bB_{GL}\bB_{LL}^{-1}\bB_{LG}$ and $ 
\bB_L= \bB_{GL}\bB_{LL}^{-1} $.

%
\subsection{Device models}\label{devices}
\subsubsection{Droop grid-forming inverter}
\begin{figure}[htbp]
	\centering \includegraphics[width=.15\columnwidth,trim={0 0 0 0},clip,angle= 90]{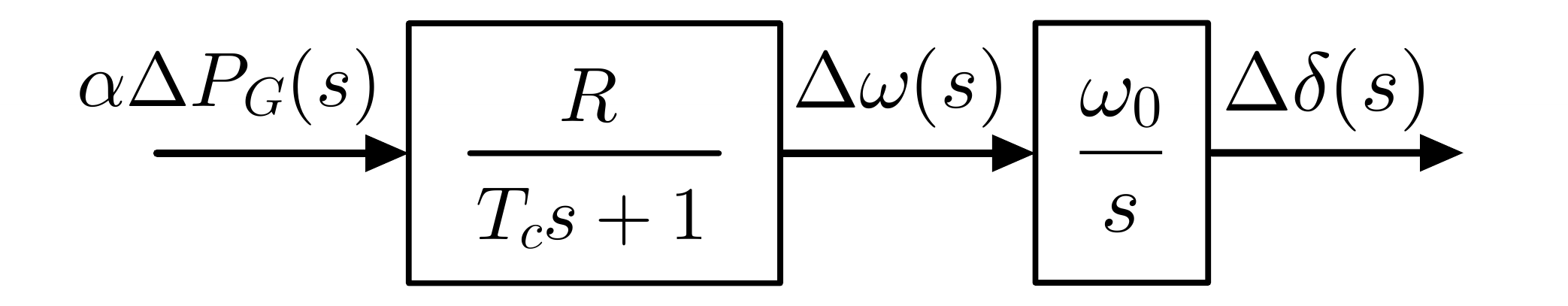}
	\caption{Low-order model of frequency response in a droop control GFM. }
	\label{fig:Pdroop}
\end{figure}

The low-order block diagram of droop control GFM frequency response is given in Fig. \ref{fig:Pdroop} \cite{sajadiSynchronizationElectricPower2022}. Power error is detected through the power measurement filter and frequency is adjusted accordingly. $R$ is the droop coefficient, $T_c$ is the time constant of the power measurement filter, and $\omega_0$ is the nominal frequency, which is $60\cdot2\pi$ in the United States. The parameter $\alpha_i$ is equal to the system base divided by the rated capacity of inverter $i$ ($\frac{S_B}{S_i}$).

In the time-domain, the GFM model defined in Fig. \ref{fig:Pdroop} can be represented by two first-order time-invariant linear ODEs shown in \eqref{ode_f_gfm}. These describe the evolution of $\Delta \delta_i$ and $\Delta \omega_i$, the voltage angle and frequency of the $i$th GFM in the network, based on the input, $\Delta P_{Gi}$.
\begin{align}\label{ode_f_gfm}
	\begin{split}
		&\frac{\rmd }{\rmd t} \Delta  {\delta}_i = \omega_0 \Delta \omega_i \\[3mm]
		&\frac{\rmd }{\rmd t} \Delta  \omega_i = -\dfrac{1}{T_{ci}}\Delta\omega_i + \dfrac{\alpha_iR_i}{T_{ci}}\Delta P_{Gi}
	\end{split}
\end{align}

Note that in this formulation, GFL buses without grid-support functionality are modeled as non-generator (static) buses. 

\subsubsection{Synchronous generator}
\begin{figure}[htbp]
	\centering \includegraphics[width=.43\columnwidth,trim={0 0 0 0},clip,angle= -90]{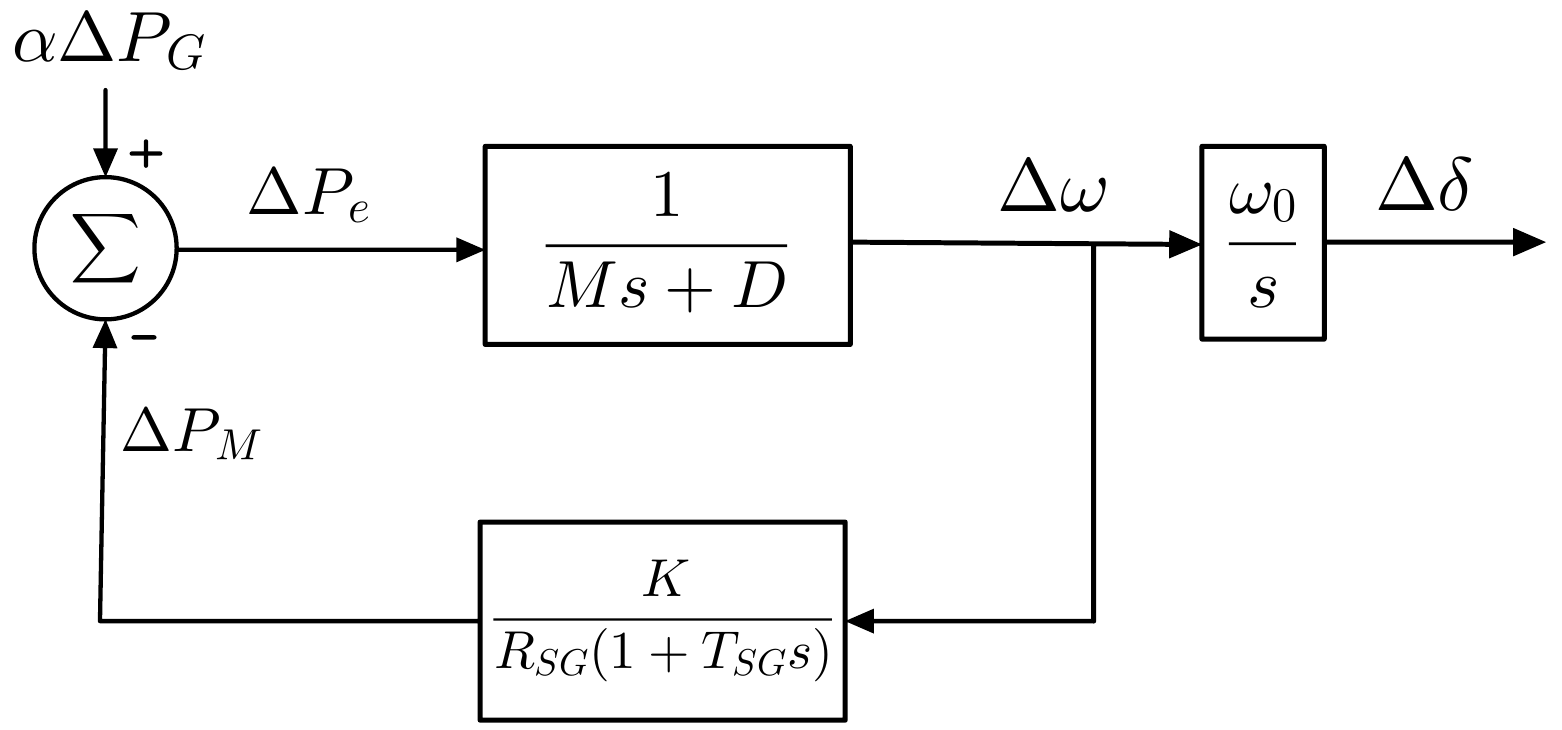}
	\caption{Low-order model of frequency response in a SG.}
	\label{fig:SG_model}
\end{figure}

The block diagram of SG frequency response is shown in Fig. \ref{fig:SG_model}. It is based on the model from \cite{sajadiSynchronizationElectricPower2022}. The shaft is represented by a first-order system with momentum $M$ and equivalent damping coefficient $D$. The turbine and governor are represented by a first order system where $K$ is the governor and turbine response gain constant, $R_{SG}$ is the droop constant, and $T_{SG}$ is the governor and turbine response time constant. To maintain the computational efficiency of the model, we assume generation devices have adequate headroom for responding to active power imbalances. 

Equation \eqref{transfer_f} is the closed loop transfer function of Fig. \ref{fig:SG_model}, with $\Delta\omega$ as the output. $T(s)$ has one zero and two poles which depend on the constant parameters $M, D, R_{SG},$ and $T_{SG}$.
\begin{align}\label{transfer_f}
	\begin{split}
		T(s) &= \frac{\Delta\omega(s)}{\alpha\Delta P_G(s)}
		\\&= \frac{sT_{SG} + 1}{s^2T_{SG}M + s(DT_{SG}+M) + (D+\frac{K}{R_{SG}})}
	\end{split}
\end{align}

In the time-domain, the SG model defined in Fig. \ref{fig:SG_model} can be represented by the three first-order time-invariant linear ODEs shown in \eqref{ode_f_sg}. These describe the evolution of the three state variables, $\Delta \delta_i, \Delta \omega_i,$ and $\Delta P_{Mi}$, for the $i$th machine in a network, where $\delta_i$ and  $\omega_i$ are the voltage angle and frequency, respectively. The change in real power seen by the $i$th generator due to the disturbance is $\Delta P_{Gi}$. 
\begin{align}\label{ode_f_sg}
	\begin{split}
		&\frac{\rmd }{\rmd t} \Delta  {\delta}_i =  \omega_0 \Delta \omega_i \\[3mm]
		&\frac{\rmd }{\rmd t} \Delta  \omega_i = - \dfrac{D_i}{M_i}\Delta \omega_i - \dfrac{1}{M_i}\Delta P_{Mi} + \dfrac{\alpha_i}{M_i}\Delta P_{Gi} \\[3mm]
		&\frac{\rmd }{\rmd t}\Delta  P_{Mi} = \dfrac{K_i}{T_{SGi}R_{SG}} \Delta \omega_i -\dfrac{1}{T_{SGi}} \Delta P_{Mi}
	\end{split}
\end{align}
\\

\subsection{Model of hybrid SG/GFM networks}\label{hybrid_Model}

To represent a network of $n$ generators in state-space form, we need to define the following vector quantities:
\begin{align*}
	&\Delta \delta \triangleq \left[\Delta \delta_{1,n} \; \hdots \; \Delta \delta_{n-1,n}\right]^\mathrm{T},  \\
	&\Delta \omega \triangleq \left[ \Delta \omega_1 \; \hdots \; \Delta \omega_n\right]^\mathrm{T} , \\
	& \Delta P_M \triangleq \left[ \Delta P_{M1} \; \hdots \; \Delta P_{Mn}\right]^\mathrm{T}, \\
	& \Delta P_L \triangleq \left[ \Delta P_{L1} \; \hdots \; \Delta P_{Ln}\right]^\mathrm{T},
\end{align*}
where $\Delta \delta_{i,n}$ denotes the relative voltage angle $\Delta \delta_{i} -  \Delta \delta_{n}$. Additionally, we let $\cG$ and $\cI$ denote the index sets associated with SG and GFM buses, respectively, and assume that the sets are mutually disjoint. For example, in a 3-node grid, if SGs reside at buses 1 and 3 and a GFM resides at bus 2, then $\cG = \{1,3\}$ and $\cI = \{2\}$. 

Recall that $\Delta P_{G}$ is a function of $\Delta\delta_G$ and $\Delta P_L$, as defined in \eqref{delta_p_gen}. Thus, concatenating \eqref{ode_f_sg} for all generators will not directly result in a state-space form network model. In \eqref{ode_f_sg}, $\Delta P_{Gi}$ is specific to the $i$th generator, but it is a function of other non-$i$th elements of $\Delta\delta_G$ and $\Delta P_L$ due to network coupling. To represent the full network in state-space form, plug \eqref{delta_p_gen} into \eqref{ode_f_sg}. The state-space input becomes $\Delta P_L$. Note that the voltage angle in \eqref{delta_p_gen} must also be converted to relative voltage angle.

The system-wide dynamics are of the form
\begin{equation}\label{state_space_f}
	\frac{\rmd}{\rmd t}
	\begin{bmatrix} 
		\Delta \delta  \\ 
		\Delta \omega \\ 
		\Delta P_M 
	\end{bmatrix} =  
	\bA_f
	\begin{bmatrix} 
		\Delta \delta \\ 
		\Delta \omega \\ 
		\Delta P_M\end{bmatrix} 
	+
	\bB_f
	\Delta P_{L} \ ,
\end{equation} 
for the block matrices
\begin{align}
	\bA_f = 
	\begin{bNiceArray}{c|c|c}[margin,baseline=c]
		\mathbf{0} & \omega_0 \bone_{-\bone} & \bzero \\
		\hline
		\tbB & \bD & \bM \\
		\hline 
		\bzero & \bK & \bT
	\end{bNiceArray} \quad \text{and} \quad 
	\bB_f = 
	\begin{bNiceArray}{c}[margin,baseline=c]
		\bzero \\
		\hline 
		\bgam \\
		\hline
		\bzero
	\end{bNiceArray}.
\end{align}
The blocks of $\bA_f$ and $\bB_f$ are as follows:
\begin{align}
	\bone_{-\bone} = 
	\begin{bmatrix}
		1 & 0 & \hdots & 0 & -1 \\
		0 & 1 & \hdots & 0 & -1 \\
		\vdots & \vdots & \ddots & \vdots & \vdots\\
		0 & 0 & \hdots & 1 & -1
	\end{bmatrix} \in \R^{(n-1) \times n}, 
\end{align}
\begin{align}
	\tbB = \text{diag}(B_i) \tbB_r \text{ for } B_i \triangleq 
	\begin{cases} 
		-\frac{\alpha_i}{M_i}, \quad & \text{ if } i \in  \mathcal{G}, \\[2mm]
		-\frac{\alpha_i R_{ci}}{T_{ci}}, \quad & \text{ if } i  \in  \mathcal{I}, 
	\end{cases}
\end{align}

where $\tbB_r$ denotes the matrix $\bB_r$ with the $n$th column removed, and 
\begin{align}
	\bD = \diag(\tilde{D}_i)  \text{ for } \tilde{D}_i \triangleq 
	\begin{cases}
		-\frac{D_i}{M_i} &\quad \text{ if } i \in \cG, \\
		-\frac{1}{T_{ci}} &\quad \text{ if } i \in \cI \ . 
	\end{cases}
\end{align}
The matrix $\bM \in \R^{n \times |\cG|}$ is constructed by taking 
\begin{align}
	\tilde{\bM} = \diag(\tilde{M_i}) \text{ for } \tilde{M_i} \triangleq
	\begin{cases}
		-\frac{1}{M_i} &\quad     \text{ if } i \in G, \\
		0 &\quad \text{ if } i \in I, 
	\end{cases}
\end{align}
and removing the zero-columns of $\tilde{\bM}$. The matrix $\bK \in \R^{|\cG| \times n}$ is given by 
\begin{align}
	\bK  = 
	\begin{bNiceArray}{c|c}[margin]
		\diag(\tilde{K}_i) & \bzero
	\end{bNiceArray} \text{ for } \tilde{K}_i \triangleq \frac{K_i}{T_{SGi}R_{SGi}}, \quad i \in \cG, 
\end{align}
while 
\begin{align}
	\bT = \diag(\tilde{T}_i) \text{ for } \tilde{T}_i \triangleq -\frac{1}{T_{SGi}}, \quad i \in \cG. 
\end{align}
Finally, 
\begin{align}
	\bgam = \diag(\tilde{\gamma}_i)\bB_L \text{ for } \tilde{\gamma}_i \triangleq
	\begin{cases}
		\frac{\alpha_i}{M_i}, &\quad \text{ if } i \in \cG, \\
		\frac{\alpha_iR_{i}}{T_{ci}}, &\quad \text{ if } i \in \cI. 
	\end{cases}
\end{align}
\\

\section{Voltage Model}\label{vmodel}
Unlike active power disturbances, which tend to have far-reaching effects due to the coupled nature of generator power angles, voltage regulation is largely a local problem, given that reactive power disturbances generally do not propagate long distances. For this reason, we approximate the voltage dynamics at each generation bus as independent from one another, as was done in \cite{trujillo2025analyticalmodelsfrequencyvoltage}. While inter-machine voltage dynamics are not modeled, the impact of the network on the dispersion of the power associate with a disturbance disturbance must be modeled to obtain accurate predictions of voltage dynamics. As such, we first discuss how a disturbance may engender changes in reactive power injection across a system. We then introduce the device-level models of voltage response and discuss how they may be combined in a single state-space representation.

\subsection{Network model}
The input change in reactive power due to the disturbance is calculated using the AC power flow method outlined in depth in \cite{trujillo2025analyticalmodelsfrequencyvoltage}. The high-level equation describing this process is \eqref{deltaQG}:
\begin{equation}\label{deltaQG}
	\Delta Q_{G,n} = Q_{G,n}(V_{post},\theta_{post}) - Q_{G0,n} ,\mspace{20mu} n \in \mathcal{D}
\end{equation}

where $Q_{G0,n}$ is the reactive power setpoint of device $n$ and $Q_{G,n}(V_{post},\theta_{post})$ is the estimated reactive power output of device $n$ after accounting for the changes in active power due to the disturbance and the change in reactive power at the disturbance bus.

\subsection{Device models}
\subsubsection{Droop grid-forming inverter}

\begin{figure}[htbp]
	\centering \includegraphics[width=.16\columnwidth,trim={0 0 0 0},clip,angle= 90]{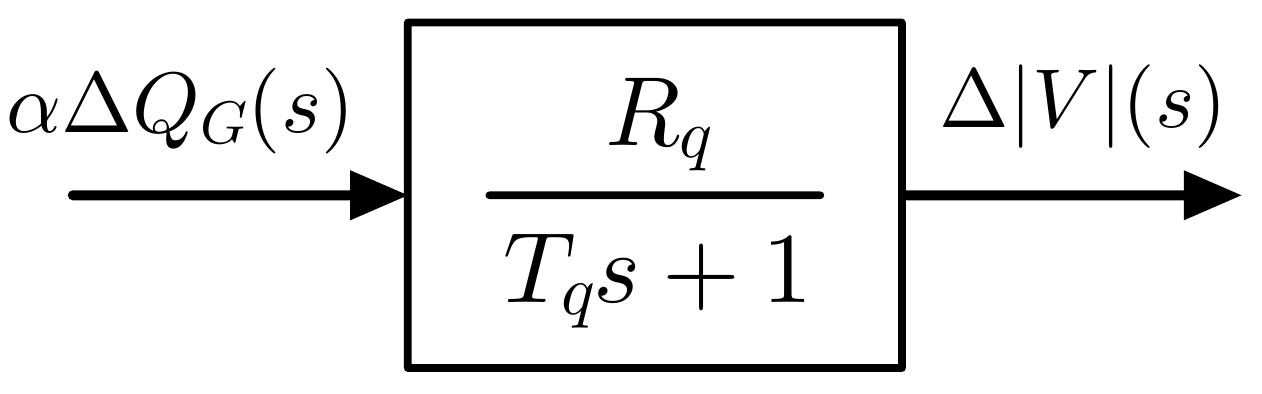}
	\caption{Low-order model of voltage response in a droop  GFM.}
	\label{fig:Qdroop}
\end{figure}

The low-order block diagram of a droop control GFM voltage response proposed in \cite{trujillo2025analyticalmodelsfrequencyvoltage} is given in Fig. \ref{fig:Qdroop}. $R_q$ is the Q-V droop coefficient, $T_q$ is the time constant of the first-order approximation of the GFM voltage step response, and $\alpha$ is equal to the system base divided by the rated capacity of the inverter.

In the time-domain, this GFM model can be represented by the single first-order time-invariant linear ODE shown in \eqref{ode_v_gfm}. This describes the evolution of $\Delta |V|_i$, the difference between the terminal voltage setpoint and the instantaneous terminal voltage the $i$th GFM in the network, based on the reactive power input disturbance, $\Delta Q_{Gi}$.

\begin{equation}\label{ode_v_gfm}
	\begin{split}
		\frac{\rmd }{\rmd t}\Delta |V|_i &= -\frac{1}{T_{qi}} \Delta|V|_i + \frac{\alpha_i R_{qi}}{T_{qi}} \Delta Q_{Gi}
	\end{split}
\end{equation}

\subsubsection{Synchronous generator}
\begin{figure}[htbp]
	\centering \includegraphics[width=.43\columnwidth,trim={0 0 0 0},clip,angle= 90]{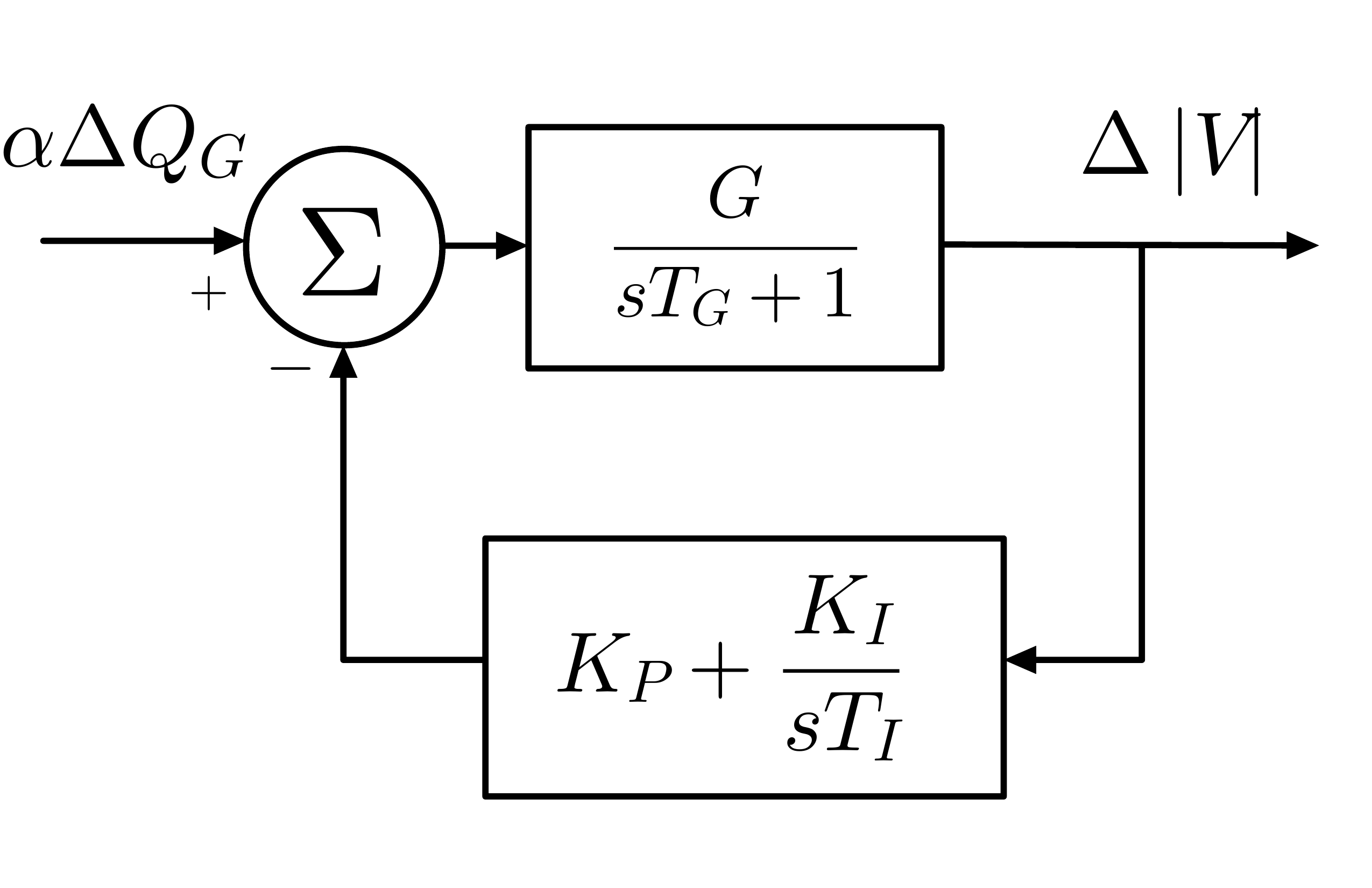}
	\caption{Low-order model of voltage response in a SG. }
	\label{fig:SGvoltage}
\end{figure}

In this paper, we present a low-order model of voltage response in a SG, which is shown in Fig. \ref{fig:SGvoltage}. As detailed in Section \ref{motivation}, AVRs create a voltage error signal to adjust the terminal voltage of a generator. However, that voltage error signal leads to a change in reactive power output by the generator, which in turn corrects any voltage deviation. This fact serves as the basis for the proposed model of voltage dynamics. 

This novel formulation was designed to be compatible with the network equation, \eqref{delta_p_gen}, and reactive power estimation procedure, \eqref{deltaQG}. In essence, we use \eqref{deltaQG} to estimate the change in reactive power output by a generator that is necessary for the generator to restore its terminal voltage to its pre-disturbance level. Once again, to preserve the computational efficiency of the model, we assume generation devices have adequate headroom for responding to reactive power imbalances. 

The closed loop transfer function of Fig. \ref{fig:SGvoltage} is shown in \eqref{transfer_v}. $T(s)$ has a zero at the origin and two poles which depend on the constant parameters $G, T_G, T_I, K_P,$ and $K_I$.
\begin{equation}\label{transfer_v}
	\begin{split}
		T(s) &= \frac{\Delta|V|(s)}{\alpha\Delta Q_G(s)}
		= \frac{Gs}{s^2T_G + s(GK_P+1) + \frac{G K_I}{T_I}}
	\end{split}
\end{equation}

In the time-domain, this SG voltage response model can be represented by two first-order time-invariant linear ODEs shown in \eqref{ode_v_sg}. Similar to \eqref{ode_v_gfm}, this describes the evolution of $\Delta |V|_i$. The second state variable, $x_{PIi}$, is the internal state of the PI block which tracks the error accumulated throughout the simulation in the $i^{th}$ SG. In this case, the error is $\Delta |V|_i$.

\begin{equation}\label{ode_v_sg}
	\begin{split}
		\frac{\rmd }{\rmd t}\Delta |V|_i &= 
		-\frac{G_iK_{Pi} + 1}{T_{Gi}}\Delta|V|_i 
		- \frac{G_iK_{Ii}}{T_{Gi}} x_{PIi} 
		+ \frac{\alpha_i G_i}{T_{Gi}}\Delta Q_{Gi} \\
		\frac{\rmd }{\rmd t}x_{PIi} &= 
		\frac{1}{T_{Ii}} \Delta|V|_i
	\end{split}
\end{equation}

The model defined by \eqref{ode_v_sg} is an linear approximation of the dynamics of a single SG. The states ($\Delta |V|_i, x_{PIi}$), input ($\Delta Q_{Gi}$), and parameters ($G_i, T_{Gi}, T_{Ii}, K_{Pi}, K_{Ii}$), are unique to each device. The devices are connected by the network, a dependency which is captured by the reactive power input, $\Delta Q_{Gi}$.

\subsection{Model of hybrid SG/GFM networks}

To represent the voltage response of a network of $n$ generators in state-space form, we need to define the following vector quantities:
\begin{align}
	\begin{split}\label{vectors_v}
		&\Delta |V| \triangleq \left[\Delta |V|_1 \; \hdots \; \Delta |V|_n\right]^\mathrm{T},  \\
		&x_{PI} \triangleq \left[ x_{PI1} \; \hdots \; x_{PIn}\right]^\mathrm{T} , \\
		& \Delta Q_G \triangleq \left[ \Delta Q_{G1} \; \hdots \; \Delta Q_{Gn}\right]^\mathrm{T} .
	\end{split}
\end{align}

We will again use $\mathcal{G}$ and $\mathcal{I}$ to denote the index sets of SG and GFM buses, respectively. Unlike the frequency response model proposed in \eqref{ode_f_sg}, the voltage response of each generator (SG or GFM) is decoupled from that of the other generators in the network. Combining \eqref{ode_v_gfm}, \eqref{ode_v_sg}, and \eqref{vectors_v}, we can form \eqref{state_space_v}, the system-wide state-space representation of all devices in the network:

\begin{equation}\label{state_space_v}
	\begin{split}
		\frac{\rmd}{\rmd t}\begin{bmatrix}
			\Delta |V| \\ x_{PI}
		\end{bmatrix} 
		&= \mathbf{A}_{v} 
		\begin{bmatrix}
			\Delta |V| \\ x_{PI}
		\end{bmatrix}
		+ \mathbf{B}_{v} 
		\begin{bmatrix}
			\Delta Q_{G}
		\end{bmatrix} , \\
	\end{split}
\end{equation}
where
\begin{align}
	\bA_v = 
	\begin{bNiceArray}{c|c}[margin,baseline=c]
		\mathbf{X} & \mathbf{Y} \\
		\hline
		\mathbf{Z}& \bzero 
	\end{bNiceArray} \quad \text{and} \quad 
	\bB_v = 
	\begin{bNiceArray}{c}[margin,baseline=c]
		\mathbf{\zeta} \\
		\hline 
		\bzero
	\end{bNiceArray} \ ,
\end{align}

\begin{alignat}{4}
	&\mathbf{X} = \text{diag}(X_i)
	&\text{ for } X_i \triangleq 
	&\begin{cases}
		-\frac{G_iK_{Pi} + 1}{T_{Gi}} & \text{ if } i \in \cG \\
		-\frac{1}{T_{qi}} & \text{ if } i \in \cI \ , 
	\end{cases}
	\\
	&\mathbf{Y} = \text{diag}(Y_i)
	&\text{ for } Y_i \triangleq 
	&\begin{cases}
		-\frac{G_iK_{Ii}}{T_{Gi}}  &\quad\   \text{ if } i \in \cG  \\
		0 &\quad\   \text{ if } i \in \cI \ ,
	\end{cases}
	\\
	&\mathbf{Z} = \text{diag}(Z_i)
	&\text{ for } Z_i \triangleq 
	&\begin{cases}
		\frac{1}{T_{Ii}}   &\qquad\quad\   \text{ if } i \in \cG  \\
		0 &\qquad\quad\   \text{ if } i \in \cI \ ,
	\end{cases}
	\\
	&\mathbf{\zeta} = \text{diag}(\zeta_i)
	&\text{ for } \zeta_i \triangleq 
	&\begin{cases}
		\frac{\alpha_i G_i}{T_{Gi}} &\qquad\ \   \text{ if } i \in \cG  \\
		0 &\qquad\ \    \text{ if } i \in \cI  \ .
	\end{cases}
\end{alignat}

\section{Parameter Estimation and Analytic Solutions}

The SG models of frequency and voltage presented in Fig. \ref{fig:SG_model} and Fig. \ref{fig:SGvoltage} are low-order approximations of SG dynamics. Some parameters for these model may be directly obtained from a manufacture's catalog, while others must be deduced from transient simulation data and by using a transfer function fitting tool, such as the \texttt{tfest} function in MATLAB \cite{TfestEstimateTransfer}. The transfer function structures should match that of \eqref{transfer_f} for the frequency model and \eqref{transfer_v} for the voltage model. 

The structure of the proposed models enables the unique benefit of finding their closed-form solutions. Namely, the models, \eqref{state_space_f} and \eqref{state_space_v}, are of the form
\begin{equation}
	\dot x(t) = \bA x(t) + \bB u, \quad x_0 = x(t_0)
\end{equation}
where $\bA$ and $\bB$ are constant matrices. The forcing, $u(t)$, is known. The solution to such a system is given by 
\begin{equation}
	x(t) = e^{\bA t}x_0 + \int_{t_0}^t e^{\bA(t - \tau)} B u(\tau) \ \rmd \tau \ ,
\end{equation}
(see \cite{ControlTheory}). In both the frequency and voltage models, a zero initial condition and constant forcing are assumed. The solutions are then 
\begin{align}
	\begin{bmatrix}
		\Delta \delta \\
		\Delta \omega \\
		\Delta P_M
	\end{bmatrix}(t) &= \int_{t_0}^t e^{\bA_f(t - \tau)}\bB_f\Delta P_L\ \rmd \tau \quad \text{ (frequency model)}, \label{freq soln}\\
	\begin{bmatrix}
		\Delta|V| \\
		x_{PI}
	\end{bmatrix}
	(t) &= \int_{t_0}^t e^{\bA_v(t - \tau)}\bB_v\Delta Q_G \ \rmd \tau \quad \text{ (voltage model)}.  \label{volt soln}
\end{align}
Further computational acceleration can be gained through the diagonalization of matrices $\bA_f$ and $\bA_v$.

\section{Validation}\label{validation}
\subsection{Frequency model}
\begin{figure}[h]
	\centering \includegraphics[width=.75\columnwidth,trim={0 0 0 0},clip]{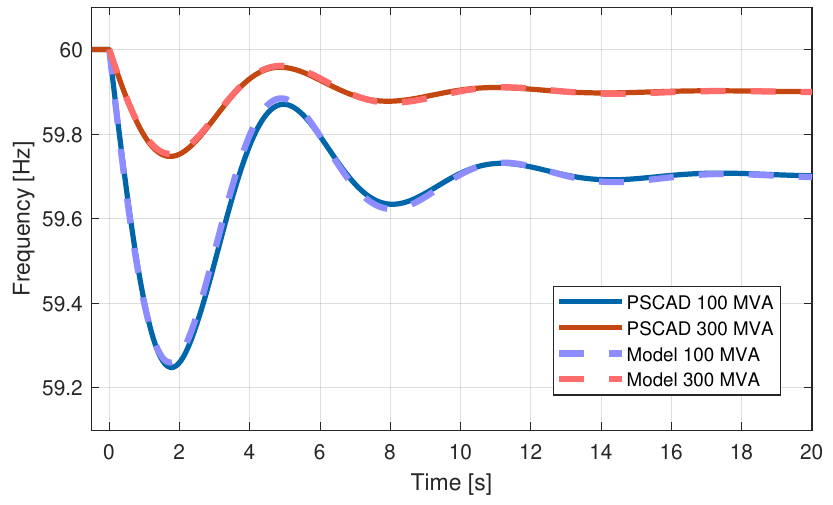}
	\caption{Frequency model parameter fitting for a single machine testbed.}
	\label{fig:onemachine_validation}
\end{figure}

A single machine testbed was built in PSCAD to validate the reduced order SG model. With some parameter fitting, there is very good agreement between the model and the full-order PSCAD simulation results (Fig. \ref{fig:onemachine_validation}). Once parameters are fitted for a SG with certain $H$, $D$, and TGOV1 parameters, the low-order model is sufficiently calibrated, even if the rated capacity of the device is changed, in which case only the $\alpha$ parameter must be updated. While here we elect to use the TGOV1 component, which was validated in \cite{wangDevelopingaPSCAD}, the frequency model parameter fitting process may be done in the same manner for SGs equipped with other turbine governor models such as GAST and IEEEG1.

\begin{figure}[htpb]
	\centering \includegraphics[width=1\columnwidth,trim={0 0 0 0},clip]{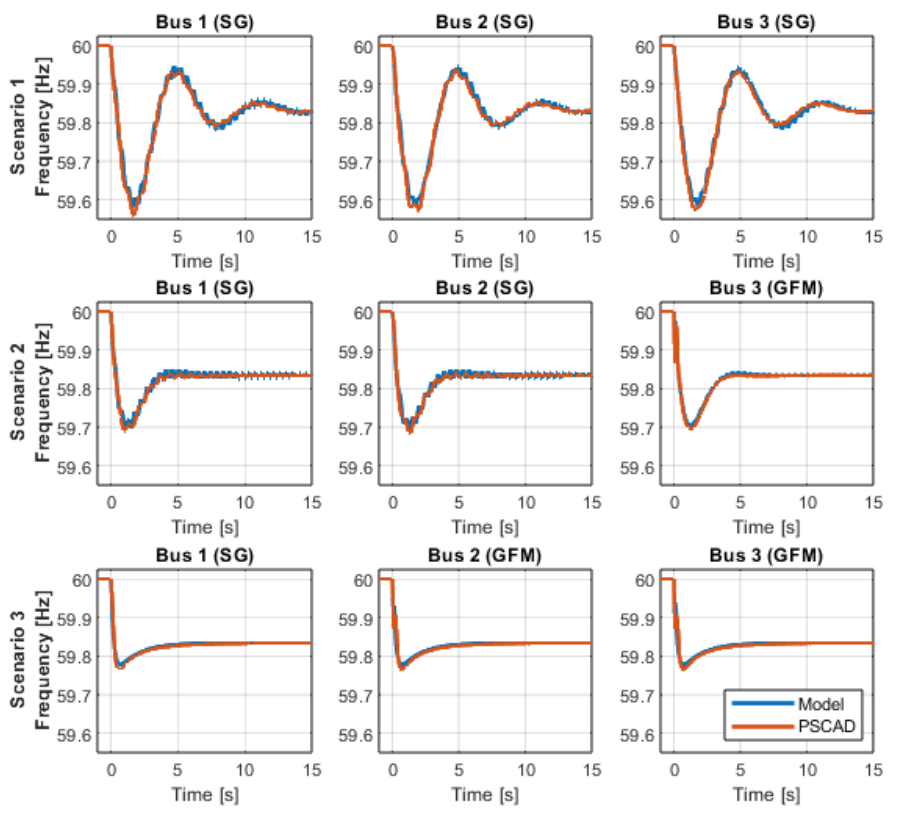}
	\caption{Validation of the proposed model against PSCAD simulations of the WECC 9 bus system.}
	\label{fig:9bus_scenarios}
\end{figure}

\begin{figure}[htpb]
	\centering \includegraphics[width=0.8\columnwidth,trim={0 0 0 0},clip]{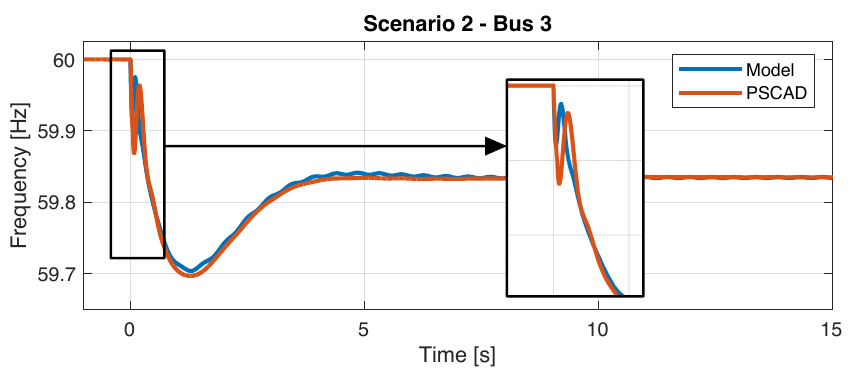}
	\caption{Frequency traces from PSCAD simulation and model prediction at Bus 3 of the WECC 9 Bus system in Scenario 2, highlighting the ability of the model to accurately capture fast and complex GFM/SG interactions.}
	\label{fig:scenario2bus3}
\end{figure}

In Fig. \ref{fig:9bus_scenarios}, we validate the model against PSCAD simulations of the WECC 9 bus system. In the first scenario, shown in the first row of  Fig. \ref{fig:9bus_scenarios}, all three generators are modeled as SGs. In the second scenario (shown in the second row), the generators at Buses 1 and 2 are modeled as SGs and the generator at Bus 3 is a droop GFM. Finally, in the third scenario, there is a SG at Bus 1 and droop GFMs at Buses 2 and 3. The PSCAD inverter models presented in \cite{kenyonOpenSourcePSCADGridFollowing2021} were utilized and the all-inverter test case was validated in \cite{trujillo2025analyticalmodelsfrequencyvoltage}. A comparison of the solve times is given in Table \ref{tab:9bus_solvetimes}. In Fig. \ref{fig:scenario2bus3}, a closer view of PSCAD and model traces from Bus 3 in Scenario 2 are shown. Here we emphasize the model's capability to accurately capture the fast dynamics introduced by GFM/SG interactions.

\begin{table}
	\centering
	\begin{tabular}{c|ccc}
		Scenario & Model [s] & PSCAD [s]& Acceleration\\ \hline
		3 SGs  (9-bus) & 0.0752 & 24.69&x328\\
		2 SGs, 1 GFM  (9-bus) & 0.0638 & 37.61&x589\\
		1 SG, 2 GFMs  (9-bus) & 0.0519 & 42.00&x809\\
		7 SG, 3 GFMs  (39-bus) & 0.1885 & 704.59& x3716\\
		5 SG, 5 GFMs  (39-bus) & 0.1749 & 1091.00& x6237\\
		3 SG, 7 GFMs  (39-bus) & 0.1577 & 1191.16&x7553\\
	\end{tabular}
	\caption{Comparison of solve times to obtain $20$ seconds of data in the WECC 9-bus and IEEE 39-bus systems.}
	\label{tab:9bus_solvetimes}
\end{table}

We next validate the model against PSCAD simulations of the IEEE 39 bus system. A load step of $307.5$ MW at Bus $15$ was simulated. In this test case, seven of the ten generation devices were droop GFMs and the remaining three were SGs. As can be seen in Fig. \ref{fig:39bus_mixed_validation}, the proposed frequency model predicts frequency response at each generation bus with high accuracy, as was the case in the $9$-bus system. The absolute value of the error in key metrics between the model and PSCAD at each bus are provided in Table \ref{tab:39freq}. In Table \ref{tab:39freq}, the Hertz-sec (HS) metric is a proxy for kinetic energy and is calculated by integrating the absolute value of frequency deviation over the transient period, as shown in \eqref{HS}. 

\begin{equation}\label{HS}
	HS = \int_{t_0}^{t_s}|f_0 - f(t)|dt \ ,
\end{equation}

where $t_0$ and $t_s$ are the onset time of transient behavior and the settling time. The error values were comparable when the proportion of GFMs to SGs was flipped (3 GFMs, 7 SGs).

\begin{figure}[htpb]
	\centering \includegraphics[width=1\columnwidth,trim={0 0 0 0},clip]{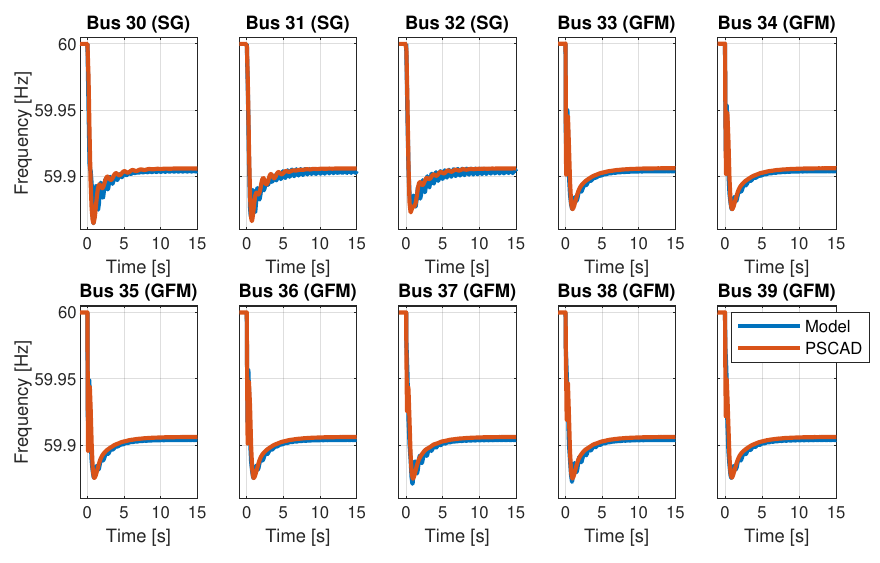}
	\caption{Validation of the proposed model against PSCAD simulation data of the IEEE 39 bus system. Seven of ten generation devices are GFMs.}
	\label{fig:39bus_mixed_validation}
\end{figure}

\begin{table}[htpb]
	\centering
	\caption{$|\text{Error}|$ of key metrics in the 39-bus system: proposed model vs. PSCAD. Simulation data is from Fig. \ref{fig:39bus_mixed_validation}.}
	\begin{tabular}{c|c|c|c}
		&\textbf{Nadir} [Hz]&\textbf{RoCoF} [Hz/s]&\textbf{HS} [Hz$\cdot$s]\\
		\hline
		Bus 30 & 0.0001 & 0.0358 & 0.0093\\
		Bus 31 & 0.0037 & 0.0117 &  0.0081 \\
		Bus 32 & 0.0021  & 0.0127&  0.0072 \\
		Bus 33 & 0.0001 & 0.0403 & 0.0082 \\
		Bus 34 & 0.0000 & 0.0442 &  0.0080 \\
		Bus 35 & 0.0001 & 0.0394&  0.0082 \\
		Bus 36 & 0.0001 & 0.0433 & 0.0081 \\
		Bus 37 & 0.0039 & 0.0238&  0.0076 \\
		Bus 38 & 0.0024 & 0.0336&   0.0078 \\
		Bus 39 & 0.0001 & 0.0298 & 0.0073 \\ \hline
		Average & 0.0013 & 0.0315 & 0.0080 \\
	\end{tabular}
	\label{tab:39freq}
\end{table}

\subsection{Voltage model}

Once again, using simulation data from a single generator testbed, the appropriate voltage model parameters were determined. As was the case with the frequency model parameters, the fitted parameters are valid for different device capacities, and only the $\alpha$ value must be changed. The model was then applied to the IEEE $39$-bus system and a load step of $307.5$ MW $+ 140.88$ MVAR at Bus $15$ was simulated. The results were compared to PSCAD simulation results, as shown in Fig. \ref{fig:39bus_mixed_validation}. The results show excellent agreement (less than 1\% error in minimum voltage in every case) between the model results and the PSCAD simulation results. The model also correctly predicted which bus (Bus $35$) would exhibit the largest voltage magnitude deviation following the disturbance. The absolute value of the error between the maximum voltage deviations predicted by the model and by PSCAD at each bus is given in Table \ref{tab:39voltages}.

\begin{figure}[htpb]
	\centering \includegraphics[width=.7\columnwidth,trim={0 0 0 0},clip]{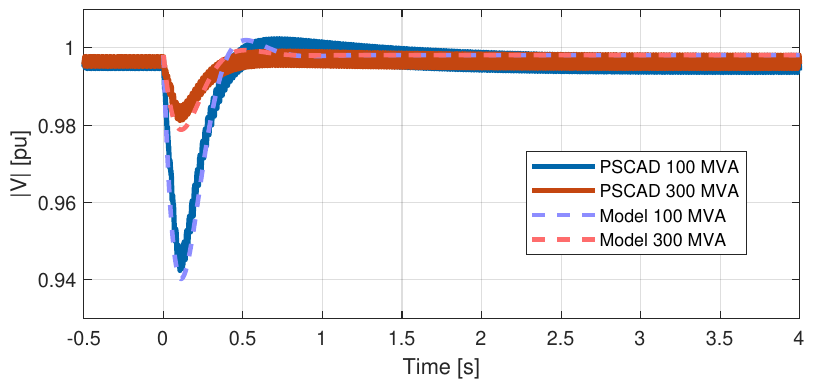}
	\caption{Voltage model parameter fitting for a single machine system.}
	\label{fig:onemachine_validation_voltage}
\end{figure}

\begin{figure}[htpb]
	\centering \includegraphics[width=1\columnwidth,trim={0 0 0 0},clip]{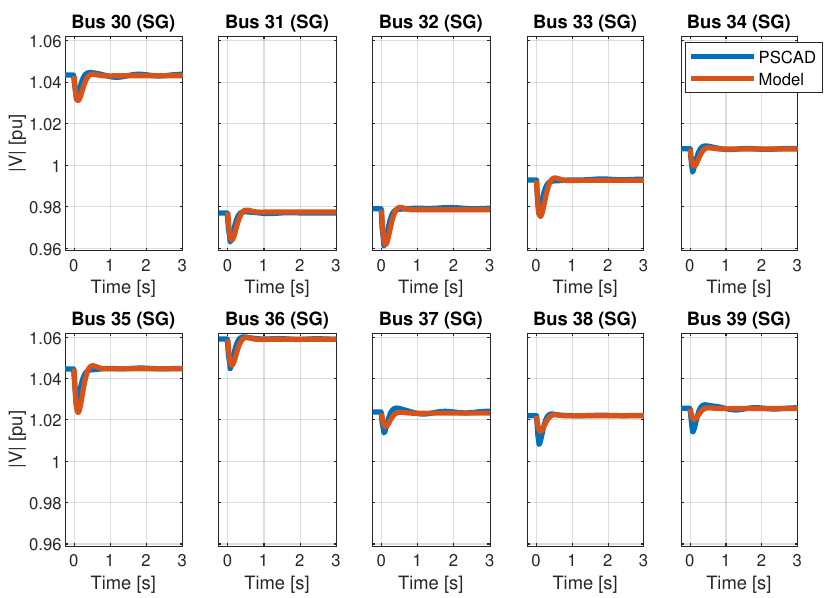}
	\caption{Validation of the proposed model against PSCAD simulation data of the IEEE 39 bus system. All devices were modeled as SGs.}
	\label{fig:39bus_mixed_validation}
\end{figure}

\begin{table}[htpb]
	\centering
	\caption{Absolute value of maximum voltage deviation error [pu] in the 39-bus system: proposed model vs. PSCAD simulations.}
	\begin{tabular}{c|c?c|c}
		&Error [pu]&&Error [s]\\
		Bus&Minimum Voltage&Bus&Time to Minimum\\
		
		\hline
		30 & 0.0013 & 30 & 0.0007\\
		31 &  0.0000 & 31 & 0.0005 \\
		32 & 0.0006  & 32&  0.0003 \\
		33 & 0.0026  &  33 & 0.0011  \\
		34 & 0.0022  & 34 & 0.0010  \\
		35 & 0.0048  & 35& 0.0009  \\
		36 & 0.0008  & 36 & 0.0008 \\
		37 & 0.0026  &  37  & 0.0005  \\
		38 & 0.0058  & 38&  0.0109  \\
		39 & 0.0056  & 39 &  0.0001  \\ \hline
		\textbf{Average} & 0.0026 & \textbf{Average}  &  0.0017\\
	\end{tabular}
	\label{tab:39voltages}
\end{table}

\section{Application and Scalability}\label{application}

After validating the accuracy of the proposed models in the previous section, we now provide an example of their scalability and include a discussion on applications.

\subsection{Scalability}
To demonstrate the scalability of the proposed models we apply them to a $2,000$-bus system  \cite{birchfieldGridStructural}. Five scenarios with increasing shares of multi-loop GFMs were simulated following a load step of $500$ MW at bus $5473$. The parameter values derived from the curve fitting shown in Figs. \ref{fig:onemachine_validation} and \ref{fig:onemachine_validation_voltage} were used. Next, the frequency and voltage trajectories at each generation bus were solved using \eqref{state_space_f} and \eqref{state_space_v}. In each scenario, SGs are supplanted by GFMs $25\%$ at a time. 

\begin{figure}[htbp]
	\centering \includegraphics[width=.85\columnwidth,trim={0 0 0 0},clip]{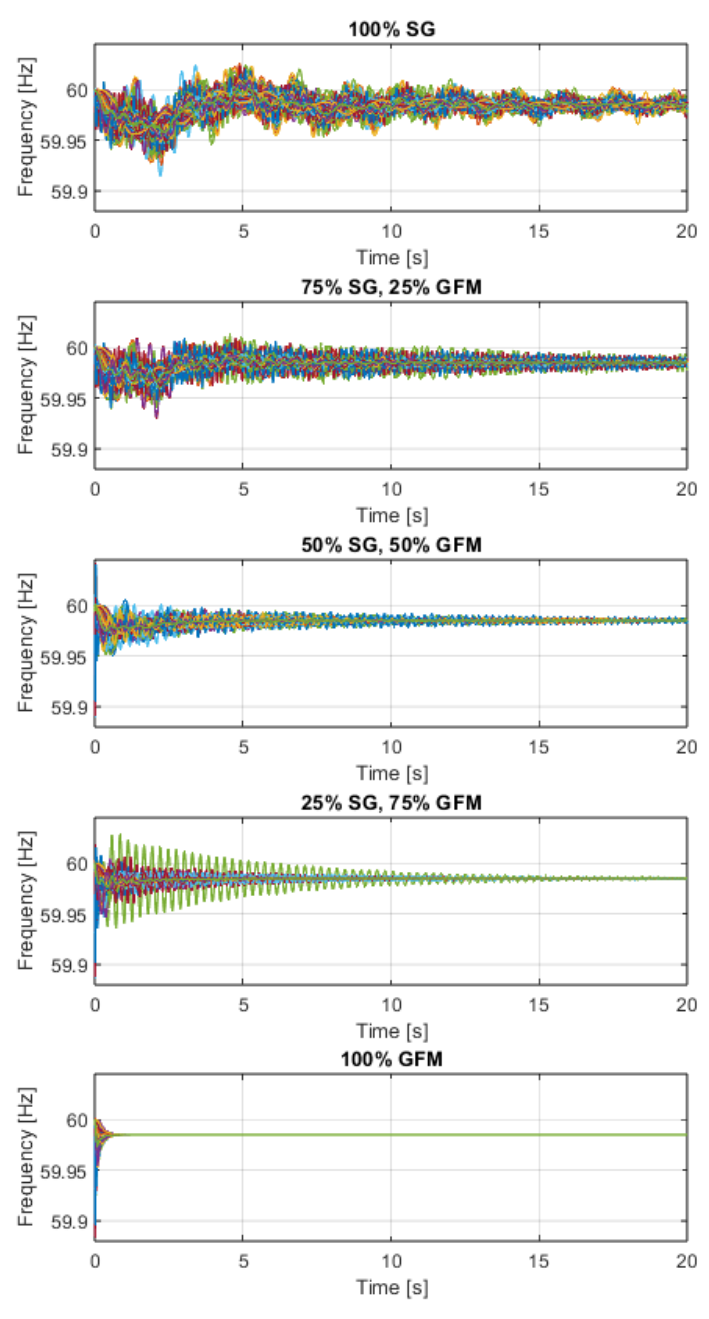}
	\caption{Frequency response across a $2,000$-bus power system in five different scenarios with increasing multi-loop droop GFM penetration.}
	\label{fig:ercot_sim}
\end{figure}

\begin{figure}[htbp]
	\centering \includegraphics[width=.97\columnwidth,trim={0 0 0 0},clip]{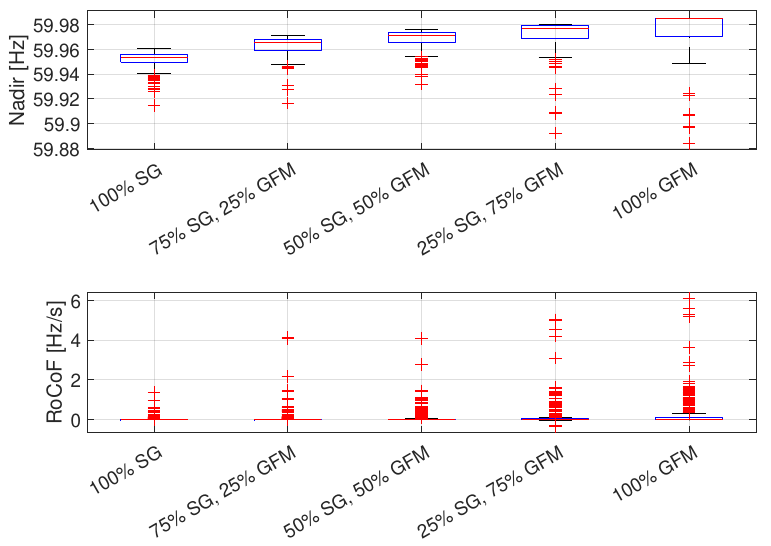}
	\caption{Box plots of nadir frequency and RoCoF across a $2,000$-bus power system in five different scenarios with increasing multi-loop droop GFM penetration.}
	\label{fig:mixed_gen_boxplots}
\end{figure}

From the trajectories shown in Fig. \ref{fig:ercot_sim} we see that as the inertia in the system decreases, interesting dynamics emerge. Whereas there is little frequency overshoot above the nominal $60$ Hz in the $100\%$ SG scenario, overshoot becomes more significant as the the amount of multi-loop GFMs increases. The exception is the $100\%$ GFM scenario, where frequency synchronizes very quickly without noticeable overshoot. These results indicate that the dynamic behavior of a system becomes more complex in low-inertia scenarios, where the frequency dynamics of SGs and IBRs are occurring on substantially different timescales. The other trend that may be observed from Fig. \ref{fig:ercot_sim} is the accelerated speed of synchronization in systems with higher shares of GFMs, with the frequencies in the all SG scenario taking $> 20$ seconds to synchronize and the frequency trajectories in the all GFM scenario synchronizing in less than a second. This more rapid synchronization is expected given that replacing SGs with GFMs is akin to replacing slow, electromechanical modes with faster, electromagnetic modes. Box plots of frequency nadir and RoCoF values across the system in each of the five scenarios are shown in Fig. \ref{fig:mixed_gen_boxplots}. A PMU sampling rate of $60$ samples/sec was assumed for calculations of RoCoF \cite{ReliabilityGuidelinePMU2016}. From these results we observe that larger instantaneous values of frequency deviation or RoCoF are not necessarily indicative of frequency instability in systems with high levels of IBRs. Interestingly, in the three scenarios where the share of GFMS was $\geq 50\%$, fast dynamics are observed immediately after the disturbance. In a conventional system comprised of SGs, these fast dynamics would be suggestive of low nadir values and likely frequency instability. However, this is clearly not the case in the depicted scenarios, where frequency deviation is arrested and settles quickly. 

\begin{figure}[htbp]
	\centering \includegraphics[width=0.75\columnwidth,trim={0 0 0 0},clip]{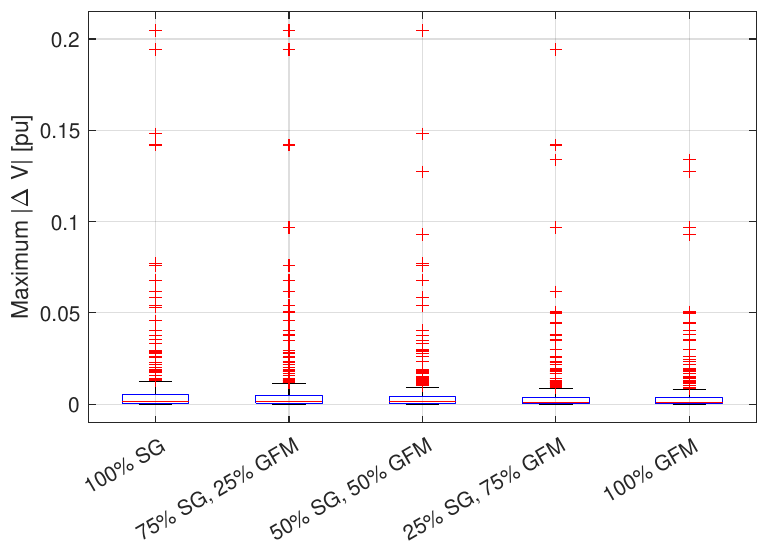}
	\caption{Box plots of maximum voltage deviation across a $2,000$-bus power system in five different scenarios with increasing multi-loop droop GFM penetration.}
	\label{fig:boxplot}
\end{figure}

For the same five scenarios depicted in Fig. \ref{fig:ercot_sim}, the proposed voltage model was also solved. At GFM buses, the voltage model of \cite{trujillo2025analyticalmodelsfrequencyvoltage} was utilized. Again, an active power load step of $500$ MW is considered. Box plots illustrating the distribution of maximum voltage deviation at each generator bus in the system for each scenario are provided in Fig. \ref{fig:boxplot}. The results showcase the importance of modeling the impact of active power disturbances on reactive power-voltage dynamics. Even following a purely active power disturbance, significant voltage deviation is seen at buses nearest the disturbance. This is because increased active power flow along transmission lines can lead to increased reactive power losses, and requires generators to inject more reactive power. It makes sense that buses with large voltage deviations are outliers considering the generally localized nature of reactive power-voltage dynamics. Still, the largest voltage deviations, which reached as high as $\approx 0.2$ pu, indicate higher ratios of estimated reactive power disturbance magnitude to device rated capacity at those buses and therefore also suggest the likelihood of voltage instability and a shortage of reactive power. Moreover, at generation buses with very large predicted voltage deviations it is important to consider the results of the proposed voltage model as a helpful starting place for further investigation into potential areas of voltage instability, that could be followed up with EMT analysis if the system is small enough. This workflow is recommended given the assumption of adequate headroom, as discussed in Section \ref{vmodel}. 

The solve times to obtain 20 seconds of frequency response data and 2 seconds of voltage response data are given in Table \ref{tab:solvetimes}. These computation times were obtained on a single AMD Ryzen $7$ $4700U$ CPU @ $2.00$GHz, $8$-core processor, with a timestep of $100 \mu$s. For scenarios with some penetration of GFMs, the voltage model solve times include the time to solve the voltage model of GFMs described in \cite{trujillo2025analyticalmodelsfrequencyvoltage}. 
\subsection{Applications}
The rapid solve times of the proposed models make them ideally suited for a variety of applications, including as a pre-EMT screening tool. When used as a screening tool, the number of computationally demanding EMT simulations that must be carried out may be reduced by first using the proposed models to identify operational scenarios most likely to result in erratic dynamic phenomena best suited for more rigorous simulation processes. This method of screening would provide a valuable alternative to standard, short-circuit ratio (SCR)-based methods which only provide scalar indicators of a power system's vulnerability to instability in a certain area \cite{GroganEMTApplications}, rather than predictions of time-domain trajectories. The models also lend themselves well for integration into several different classes of power system optimization problems including transient stability constrained unit commitment and economic dispatch as well as planing problems like optimizing the location of grid-forming assets. Finally, future work should explore leveraging the proposed models for co-simulation with other simulation tools, like EMT software, to model large grid systems outside a primary area of interest.

\begin{table}
	\centering
	\begin{tabular}{c|c|c}
		& Solve Time [s] & Solve Time [s]\\
		Scenario & Frequency Model & Voltage Model\\
		\hline
		100\% SG  & 8.5219 & 0.5064\\
		75\% SG, 25\% GFM & 7.7450 &0.3161\\
		50\% SG, 50\% GFM & 6.3312 & 0.1867\\
		25\% SG, 75\% GFM & 5.6051 & 0.1032\\
		100\% GFM  & 4.7605 & 0.0542\\
	\end{tabular}
	\caption{Solve times of the proposed models to obtain $20$ seconds of frequency response data and $2$ seconds of voltage response data under five different GFM penetration scenarios. Simulations are of a $2,000$-bus system \cite{birchfieldGridStructural}.}
	\label{tab:solvetimes}
\end{table}

\section{Conclusion}

As decarbonization goals drive the rapid deployment of IBRs, the ability to accurately and efficiently simulate the dynamics - \textit{both} frequency and voltage - of these low-inertia system becomes increasingly important. In this paper, we present low-order models of frequency and voltage response for networks comprised of both SGs and IBRs. The models were extensively validated against simulations of the WECC 9 bus system and the IEEE 39 bus system under different IBR-penetration scenarios. Then, the models were applied to a 2,000 bus system to demonstrate their scalability. The proposed models are dramatically more efficient than EMT analysis provided certain nonlinearities, such as saturation dynamics, can be neglected. As such, the models are well-suited to be embedded within a variety of power systems planning and operations problems.

\section{Acknowledgments}\label{acknowledgments}
 We would like to thank Dr. Mostafa Sedighizadeh, Harvey Scribner, and Nick Parker at Southwest Power Pool (SPP), Dr. Mohammadi Mohammadi of the Australian Energy Market Operator (AEMO), and Dr. Bruno Leonardi of the New York Independent System Operator (NYISO) for their insightful feedback and constructive comments. Additional thanks to Matthew Baughman and Fiona Majeau of University of Colorado Boulder for their insightful comments. This material is based upon work supported by the National Science Foundation Graduate Research Fellowship under Grant No. DGE $2040434$ and in part by the National Science Foundation’s Advancing Sustainability through Power Infrastructure for Roadway Electrification (ASPIRE) Engineering Research Center and National Science Foundation, United States Award $1941524$.

This work was authored in part by the National Renewable Energy Laboratory, operated by Alliance for Sustainable Energy, LLC, for the U.S. Department of Energy (DOE) under Contract No. DE-AC$36$-$08$G$O28308$. The views expressed in the article do not necessarily represent the views of the DOE or the U.S. Government. The U.S. Government retains and the publisher, by accepting the article for publication, acknowledges that the U.S. Government retains a nonexclusive, paid-up, irrevocable, worldwide license to publish or reproduce the published form of this work, or allow others to do so, for U.S. Government purposes.


\end{document}